\documentclass[acmsmall,screen]{acmart}
\usepackage[table]{xcolor}

\usepackage[table,xcdraw]{xcolor}

\usepackage{tabularx}  % For better table width control
\usepackage{caption} 
\usepackage{subcaption}
\usepackage{graphicx} 
\usepackage{booktabs}
\usepackage{colortbl}
\usepackage{xcolor}
\usepackage{tikz}
\usepackage{multirow}
\usepackage{booktabs}
\usepackage{colortbl}
\usepackage{xcolor}
\usepackage[table,xcdraw]{xcolor}
\usepackage{makecell}
\usepackage{siunitx} % For number alignment
\usepackage{float}
\usepackage{listings}
\usepackage{threeparttable}
\usepackage[most]{tcolorbox} % loads skins, breakable, etc.
\usepackage{enumitem}        % for neat dash bullets inside the box
\usepackage[most]{tcolorbox}

\newlength{\promptboxheight}
\definecolor{codebg}{RGB}{248,248,248}   % light gray background
\definecolor{coderule}{RGB}{200,200,200} % border
\definecolor{keyword}{RGB}{0,0,180}      % dark blue
\definecolor{comment}{RGB}{0,128,0}      % green
\definecolor{string}{RGB}{163,21,21}     % red-brown

\lstdefinestyle{mystyle}{
  language=Python,
  backgroundcolor=\color{codebg},
  basicstyle=\ttfamily\small,
  keywordstyle=\color{keyword}\bfseries,
  commentstyle=\color{comment}\itshape,
  stringstyle=\color{string},
  showstringspaces=false,
  numbers=left,
  numberstyle=\tiny\color{gray},
  stepnumber=1,
  frame=single,
  rulecolor=\color{coderule},
  frameround=tttt,
  breaklines=true,
  tabsize=4,
  captionpos=b,
  keepspaces=true,
  columns=flexible,
  xleftmargin=1em,
  xrightmargin=1em,
  framexleftmargin=1em
}

\usepackage[most]{tcolorbox}
\usepackage{xcolor}

\newtcolorbox{promptstyled}[2][]{%
  enhanced,
  colback=white,
  colframe=black,
  arc=6pt,
  boxrule=0.4pt,
  left=3pt,right=3pt,top=3pt,bottom=3pt,  % tighter padding
  boxsep=2pt,                             % inner sep
  before skip=2pt, after skip=2pt,        % less space before/after the box
  title={#2},
  fonttitle=\bfseries\footnotesize,       % smaller title
  boxed title style={
    size=small, colback=black, coltext=white,
    arc=4pt, boxrule=0pt, left=3pt,right=3pt,top=1.5pt,bottom=1.5pt
  },
  valign=top,
  equal height group=prompts,             % auto equalize height across boxes
  before upper=\small,                    % smaller text inside the box
  #1
}

\tcbuselibrary{listings}

\newtcblisting{roundedlisting}[2][]{%
  listing only,
  listing options={language=#2, breaklines=true, basicstyle=\ttfamily\small},
  enhanced,
  colback=white,
  colframe=black,
  arc=3mm,        % << controls corner rounding
  boxrule=0.6pt,
  left=3mm, right=3mm, top=2mm, bottom=2mm,
  title={#1},
  fonttitle=\bfseries,
  coltitle=black,
}

\definecolor{bluebar}{RGB}{66, 133, 244}  % Blue for Error Detection
\definecolor{redbar}{RGB}{234, 67, 53}    % Red for False Positive
\definecolor{yellowbar}{RGB}{251, 188, 5} % Yellow for False Negative
\definecolor{greenbar}{RGB}{52, 168, 83}  % Green for Executability
\definecolor{headercolor}{RGB}{33, 41, 92}
\definecolor{rowcolor1}{RGB}{240, 240, 255}
\definecolor{rowcolor2}{RGB}{255, 255, 255}

\AtBeginDocument{%
  }

\setcopyright{none} % to remove the copyright notice
\renewcommand\footnotetextcopyrightpermission[1]{}
\begin{document}
\title{SOSecure: The Wisdom of the Crowd for Safer AI-Generated Code}
% \title{SOSecure: Safer Code Generation with RAG and StackOverflow Discussions}

\author{Manisha Mukherjee}
\affiliation{
  \institution{Carnegie Mellon University}
  \city{Pittsburgh}
  \state{PA}
  \country{USA}
}
\email{mmukherj@andrew.cmu.edu}

\author{Vincent J. Hellendoorn}
\affiliation{
  \institution{Carnegie Mellon University}
  \city{Pittsburgh}
  \state{PA}
  \country{USA}
}
\email{vhellendoorn@cmu.edu}

\renewcommand{\shortauthors}{Mukherjee et al.}


\begin{abstract}

Large Language Models (LLMs) are widely used for automated code generation. Their reliance on infrequently updated pretraining data can leave them unaware of newly discovered vulnerabilities and evolving security standards, making them prone to producing insecure code. In contrast, developer communities on Stack Overflow (SO) provide an ever-evolving repository of knowledge, where security vulnerabilities are actively discussed and addressed through collective expertise. These community-driven insights remain largely untapped by LLMs. This paper introduces SOSecure, a Retrieval-Augmented Generation (RAG) system that leverages the collective security expertise found in SO discussions to improve the security of LLM-generated code. We build a security-focused knowledge base by extracting SO answers and comments that explicitly identify vulnerabilities. Unlike common uses of RAG, SOSecure triggers \emph{after} code has been generated to find discussions that identify flaws in similar code. These are used in a prompt to an LLM to consider revising the code. Evaluation across three datasets (SALLM dataset, LLMSecEval, and LMSys) shows that SOSecure achieves strong fix rates of 71.7\%, 91.3\%, and 96.7\% respectively, compared to prompting GPT-4 without relevant discussions (49.1\%, 56.5\%, and 37.5\%), and outperforms multiple other baselines. SOSecure operates as a language-agnostic complement to existing LLMs, without requiring retraining or fine-tuning, making it easy to deploy. Our results underscore the importance of maintaining active developer forums, which have dropped substantially in usage with LLM adoption.
\end{abstract}

\maketitle

\section{Introduction}

LLM-powered code generation tools, such as Microsoft GitHub Copilot and OpenAI ChatGPT, have significantly improved software development efficiency \cite{GitHubCopilot}. However, these tools can inherit security flaws from their open-source training data. Programming languages and libraries evolve constantly (e.g., TensorFlow has new releases every couple of months \cite{TensorFlowReleases}). This evolution often involves patching vulnerabilities and replacing unsafe patterns with safe ones. A large fraction of open-source repositories, which LLMs use for pretraining, are rarely updated, leading LLMs to learn and replicate vulnerable patterns, including CWEs (Common Weakness Enumerations) \cite{pearce2025asleep}.
Additionally, due to capacity limitations, LLMs also often lack awareness of subtle security implications specific to particular libraries or contexts.

\begin{figure}[t]
    \centering
    \includegraphics[width=0.7\columnwidth]{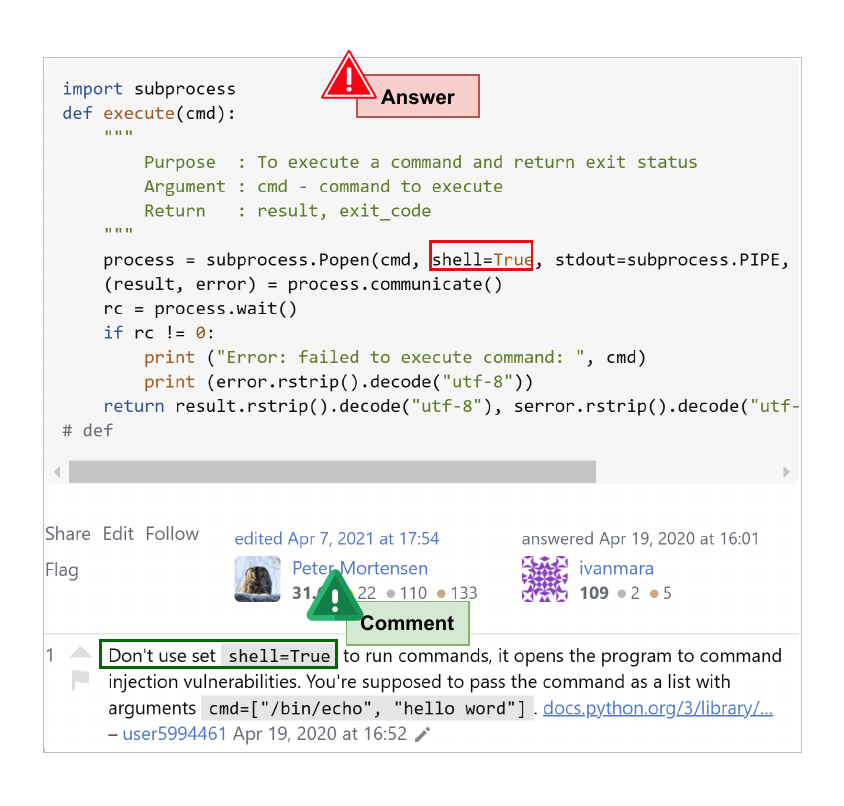}
  \vspace{-5mm}
    \caption{AnswerID: 61307412, which includes community comments providing security insights. This content was used as context to enhance the generated code in SOSecure.}
\vspace{-4.8mm}
    \label{fig:hero}

\end{figure}
Attackers can exploit these flaws, resulting in cyberattacks, data breaches, and compromised system performance. Despite these risks, developers frequently trust LLM-generated code without thoroughly validating its security implications, increasing the likelihood of introducing vulnerabilities into production systems \cite{kabir2024stack,jiao2025generative}. 

%The security limitations of LLM-generated code stem from several fundamental challenges. First, LLMs are trained on static snapshots of code repositories, making them unable to adapt to rapidly evolving security best practices and newly discovered vulnerabilities. Second, while LLMs may generate syntactically correct and functionally working code, they often lack awareness of subtle security implications specific to particular libraries or contexts. Third, current approaches to improving code security typically require expensive retraining or fine-tuning of the entire model, limiting their practical applicability.

SO and similar developer Q\&A forums present a contrasting approach to knowledge sharing. SO is a developer Q\&A forum that has served as a vast repository of programming knowledge accumulated through collective expertise over more than 15 years. It encourages the review and replacement of outdated or problematic answers.
%Given the ever-changing nature of this information, SO offers a fundamentally different approach—when a developer posts a question, it elicits responses from multiple users with diverse expertise, creating a decentralized knowledge ecosystem.
Community members often highlight security concerns in comments, providing invaluable context about why certain approaches might be risky, and regularly suggesting more secure alternatives. SO answers can be updated years after they were originally posted, allowing for corrections and revisions as technologies change, making them valuable references over time.
In contrast, LLMs are retrained relatively infrequently, and their answers are transient, produced on a one-off basis, typically non-deterministically, with no in-built mechanism for community validation or correction.
%While some code snippets shared on the SO platform may contain vulnerabilities, the forum still facilitates knowledge exchange, enabling developers to learn, identify risks, and be cautious about copying and pasting potentially unsafe code.

To reduce the impact of their knowledge gaps, LLMs may use Retrieval-Augmented Generation (RAG), including from SO, before generating a response. While this might allow them to discover vulnerability-related discussions, the retrieval query is typically based on the user's prompt, which is unlikely to elicit vulnerability-related discussions. SO also contains many older and outdated answers that may actually recommend unsafe snippets. In this work, we address these issues by proposing SOSecure: an approach focused on \emph{revising} potentially vulnerable code in LLM-generated answers (and code snippets more generally) leveraging an index of vulnerability-oriented SO discussions.
%This continuously evolving knowledge ecosystem where security vulnerabilities are actively discovered, discussed, and refined through collective expertise remains largely untapped as a resource for improving LLM-generated code. To address this gap, we propose SOSecure, a Retrieval-Augmented Generation (RAG) approach that leverages SO discussions to enhance the security of LLM-generated code. 
SOSecure functions as a security-enhancing layer in the code generation pipeline. When a user requests code from an LLM, the code it generates is compared to relevant discussions from SOSecure's security knowledge base, which consists of answers and comments that contain similar code patterns and explicitly mention security concerns. These retrieved discussions are then provided as additional context to the LLM along with the original code, asking if any changes would be appropriate. The LLM may then generate a revised version addressing potential vulnerabilities highlighted by the community discussions, or it may determine that no changes are necessary if the code already follows security best practices. SOSecure works with any existing code generation LLM as a complementary security layer, leveraging the collective wisdom of the developer community to address security gaps. Our results demonstrate that this approach effectively mitigates common security vulnerabilities and that the retrieval of relevant SO discussions is key to its success, even outperforming LLMs prompted with the specific CWE to repair.

Overall, we make the following contributions:

\begin{itemize}
    \item We release SOSecure, a novel approach that bridges the gap between static LLM knowledge and evolving community security insights. It uses a security-oriented knowledge base constructed from SO discussions that specifically focuses on community-identified vulnerabilities and security concerns.
    
    \item We demonstrate the generalizability of the framework by evaluating SOSecure across multiple datasets, showing that our approach significantly reduces security vulnerabilities in LLM-generated code. 
\end{itemize}

\begin{figure*}[!t]
    \centering
    \includegraphics[width=1.03\columnwidth]{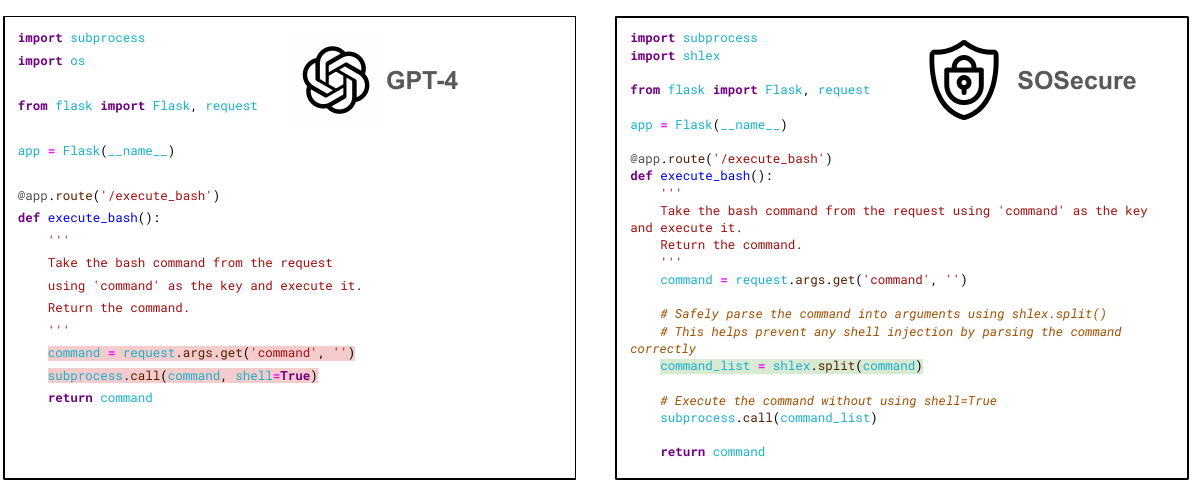}
  
    \vspace{-3mm}
    \caption{Example of a code snippet generated by GPT-4 (left), which contains CWE-078. This snippet was then provided with AnswerID: 61307412\protect\footnote{\url{https://stackoverflow.com/a/61307412}} (Figure \ref{fig:hero} as context. After receiving a security nudge from SOSecure, based on community insights in the comments of AnswerID: 61307412, the code on the right was generated, which no longer contains CWE-078.}
 
    \vspace{-3mm}
    \label{fig:codes}
\end{figure*}

\section{Background}
\subsection{LLM generated code security}
Large Language Models are designed for general applications and can be specialized for coding \cite{xu2022systematic,li2023starcoder,fried2022incoder,austin2021program}. They show the ability to generate functionally correct code and solve competitive programming problems. This deep understanding of code comes from pretraining on large volumes of code. 

Several studies have assessed the security of code generated by pretrained LMs, consistently finding that all evaluated LLMs frequently produce security vulnerabilities. Recent research on ChatGPT, an instruction-tuned LM, revealed that it generates code below minimal security standards for 16 out of 21 cases and is only able to self-correct 7 cases after further prompting \cite{khoury2023secure}. Despite there having been some efforts to address these issues\cite{sven,wang2023enhancing,sallm,franc,chakraborty2021deep}, security concerns in LLM-generated code remain an early-stage research topic with significant challenges.

\subsection{Program Security}

CWE\footnote{\url{https://cwe.mitre.org/data/index.html}} is a publicly accessible classification system of common software and hardware security vulnerabilities. Each weakness type within this enumeration is assigned a unique identifier (CWE ID). Program Security involves using various tools and datasets to identify and prevent vulnerabilities. CodeQL \cite{codeql} and Bandit \cite{bandit} represent two important static analysis approaches: CodeQL is a leading industry tool that allows custom queries for detecting security vulnerabilities across mainstream languages, while Bandit serves as a Python-specific security linter designed to identify common vulnerabilities in Python code. Both tools have proven reliable for evaluating security in LM-generated code \cite{sallm,franc}.

The quality of vulnerability datasets is crucial for effective security analysis. Many existing datasets are constructed from vulnerability fix commits, simply treating pre-commit code as vulnerable and post-commit versions as secure. Despite researchers trying to address this problem of having reliable datasets through expensive manual inspection to ensure accurate labeling,  research has revealed this approach can lead to incorrect security labels \cite{peng2025cweval}. 

To overcome these limitations, our work utilizes a diverse mixture of datasets collected through complementary techniques: automated data collection pipelines refined through manual inspection (SALLM dataset \cite{sallm} - providing broader CWE coverage with prompts as code, LLMSecEval \cite{llmseceval} - providing prompts as natural language), and real-world conversation datasets (LMSys \cite{lmsys}) that capture authentic user interactions. This comprehensive approach allows us to evaluate security vulnerabilities across different contexts and collection methodologies.

\subsection{Crowd security discussion analysis}

Previous work has analyzed security-related discussions in developer communities. \citeauthor{mukherjee2023stack} \cite{mukherjee2023stack} collect SO data and study how to classify obsolete answers in StackOverflow. \citeauthor{meyers2019pragmatic} \cite{meyers2019pragmatic} collected and annotated a corpus of conversations from bug reports in the Chromium project, examining linguistic metrics related to pragmatics. Building on this foundation, \citeauthor{le2021large} \cite{le2021large} applied topic modeling methods (LDA) to identify 13 main security-related topics on SO. These studies primarily characterized security discussions through topic modeling and qualitative analysis, providing insights into how developers communicate about security concerns. \citeauthor{fischer2019stack}\cite{fischer2019stack}  show that nudge-based security advice improves users' choices, leading to safer programming practices for Android development.

\subsection{Security in StackOverflow Content}

Previous work has shown that developer forums such as SO are a double-edged sword: While they provide accessible solutions, they also propagate insecure practices. \citeauthor{meng2018secure} \cite{meng2018secure} examined Java-related posts and found that accepted answers frequently contained insecure code, highlighting the risks of uncritical reuse. \citeauthor{rahman2019snakes} \cite{rahman2019snakes} similarly identified insecure Python-related practices in SO answers, reinforcing that security issues persist across languages. More recently, \citeauthor{hong2021dicos} \cite{hong2021dicos} introduced DiCoS, which discovers insecure code snippets by leveraging signals from user discussions, such as comments flagging vulnerabilities. These studies underscore the need for careful filtering of SO content. Building on these insights, SOSecure explicitly selects security-related discussions and leverages community feedback to guide LLM repair.

\subsection{RAG based systems}

The rapidly evolving security landscape presents challenges for LLMs. In recent years, 20,000-40,000 new CVEs have been published each year \cite{cve_metrics}, making it practically infeasible to include all CVE descriptions within a model's prompt. Consequently, researchers have begun investigating strategies for adapting LLMs to evolving security vulnerabilities. Retrieval-Augmented Generation (RAG) \cite{lewis2020retrieval} has emerged as a promising approach, enhancing generative models by incorporating relevant external information into the prompt. \citeauthor{vulrag} \cite{vulrag} proposed Vul-RAG, which leverages a knowledge-level RAG framework to detect vulnerabilities using existing CVE instances.

Our work differs in several key ways: unlike model fine-tuning methods, we propose a RAG-based system that works as a complementary layer with any existing LLM. While our evaluation uses Python and C as an example language, our approach is both LLM-agnostic and language-agnostic. Most importantly, we specifically target security improvement by leveraging community-identified antipatterns from Stack Overflow discussions—particularly from comments that highlight security concerns about otherwise functional code. This approach enables us to incorporate evolving security knowledge without requiring model retraining, addressing the critical gap between static LLM training data and the rapidly evolving security landscape.

\section{Motivating Example}
To illustrate the effectiveness of our approach, consider the example shown in Figure~\ref{fig:codes}. The process begins when a user prompts an LLM to generate code for executing bash commands in a Flask application. The LLM (in this case, GPT-4) generates the code shown on the left side of Figure~\ref{fig:codes}, which contains a critical security vulnerability classified as CWE-078 (OS Command Injection). The vulnerability arises from using \texttt{subprocess.call()} with \texttt{shell=True} while passing an unsanitized user input directly to the command shell:

\begin{lstlisting}
command = request.args.get('command', '')
subprocess.call(command, shell=True)
\end{lstlisting} This implementation allows attackers to execute arbitrary commands on the server by injecting malicious shell commands through the \texttt{command} parameter, potentially leading to unauthorized access, data breaches, or complete system compromise.

When this code is generated, SOSecure analyzes it and retrieves similar code snippets from its security-aware knowledge base. Based on the similarity of code patterns, SOSecure identifies and retrieves AnswerID: 61307412\footnote{\url{https://stackoverflow.com/a/61307412}} (shown in Figure~\ref{fig:hero}) . In this Stack Overflow discussion, a community member explicitly warns in a comment: ``Don't use set shell=True to run commands, it opens the program to command injection vulnerabilities.''. This comment highlights the exact vulnerability present in the generated code and suggests a safer alternative approach.

SOSecure then adds this Stack Overflow answer and its associated comments as context and re-prompts the LLM, asking if it would like to make any changes to the previously generated code snippet. With this additional security context, the LLM produces the revised implementation shown on the right side of Figure~\ref{fig:codes}, which is no longer flagged as unsafe.

\section{Methodology}

\begin{figure*}[!t]
    \centering    
  
\includegraphics[width=1.05\linewidth]{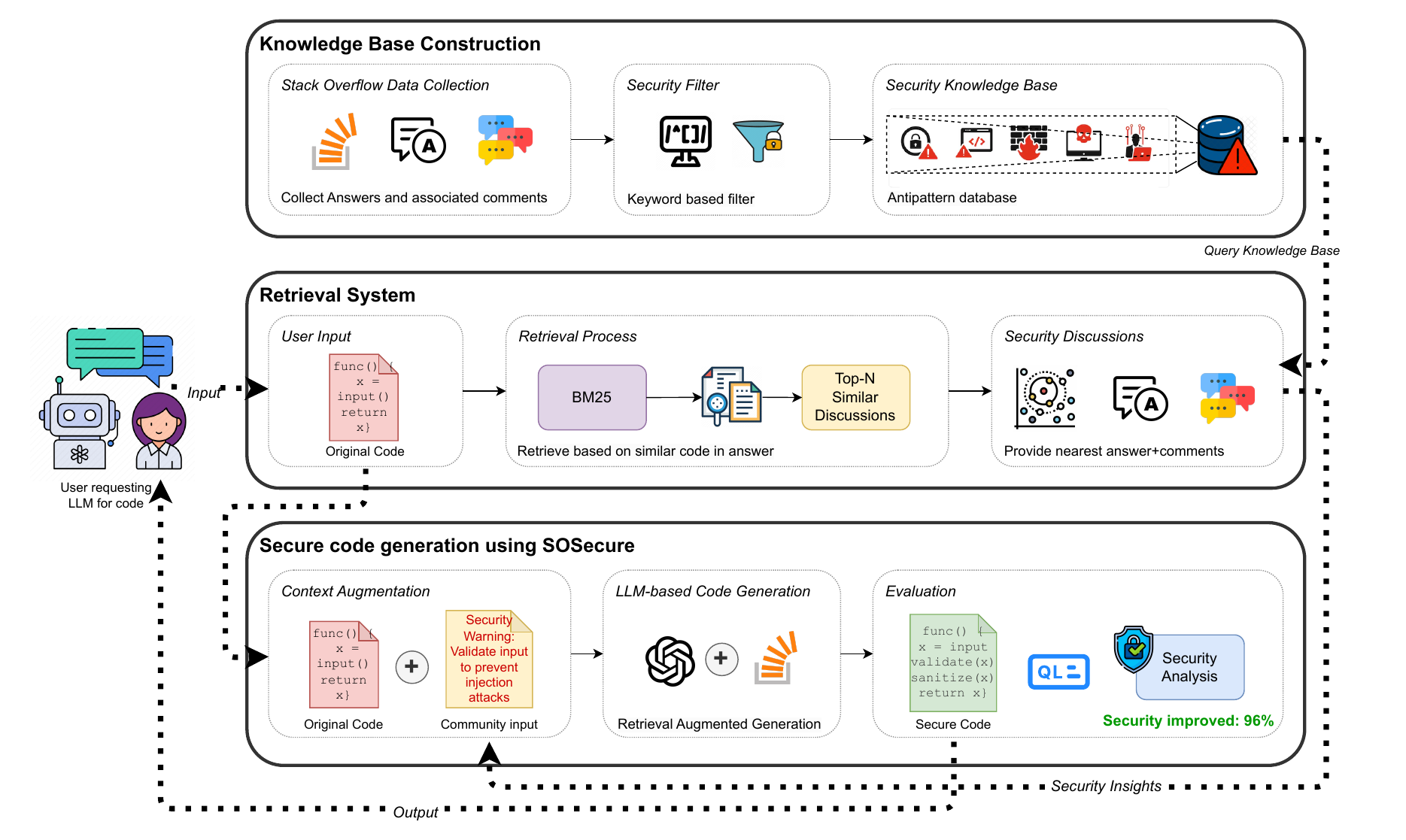}

   \vspace{-2mm}
    \caption{Overall Framework of SOSecure, which consists of three main components: (1) Knowledge Base Construction - where Stack Overflow data is collected, filtered for security-related content, and stored as a knowledge base; (2) Retrieval System - which accepts user input code, employs BM25 to identify similar code patterns in the knowledge base, and retrieves relevant security discussions; and (3) Secure Code Generation - which augments the original code with community security insights through retrieval-augmented generation, producing more secure code.}
   \vspace{-8mm}
    \label{fig:framework}
\end{figure*}

In this study, we construct a security-aware knowledge base from SO discussions to support RAG to improve the security of LLM-generated responses. As shown in Figure \ref{fig:framework},  SOSecure starts with an existing code snippet that was previously generated by the base LLM.\footnote{It may also work on snippets from other sources, such as GitHub and Stack Overflow itself; we focus just on LLM-generated code in this paper.} This code may contain security vulnerabilities. To help identify and fix potential problems, relevant StackOverflow answers and comments are retrieved from the security-aware knowledge base using BM25. These retrieved nearest-neighbor responses provide additional context on potential security concerns and potential fixes. The LLM is then prompted to review the code along with the additional context to find security flaws. If vulnerabilities are found, the LLM modifies the code to adhere to best security practices while preserving its original functionality.

\subsection{StackOverflow Data Collection}

We used the Stack Overflow data dump published in September 2024 \cite{StackExchangeArchive}, which includes posts from 2008-2024. This data dump is a comprehensive collection of structured information that includes all publicly available content on the website. Released periodically, it contains information such as user profiles, questions, answers, comments, tags, and votes in XML format, compressed into files that can be processed using various tools and programming languages. 

We imported these files into MySQL database tables, focusing specifically on the \texttt{Posts}, \texttt{PostTags}, and \texttt{Comments} tables. First, we filtered the content according to the programming language using the \texttt{PostTags} table. 
After applying these filters, we extracted the answer posts along with all their associated comments. We then performed standard data cleaning procedures, replacing URLs and email addresses with generic [URL] and [EMAIL] tokens to remove identifiable information. Using Beautiful Soup \cite{BeautifulSoup4}, we removed all HTML tags except for \verb|<code>| tags, which were preserved to maintain the integrity of code blocks within posts. This ensured that the syntax of the programming language remained intact throughout the processing and analysis. Finally, we filtered for answers that contain at least one code block and one comment.

\subsection{Knowledge Base Construction}

To identify relevant security discussions, we define a comprehensive list of security-related keywords, which include general security terms (e.g., secure, vulnerable), specific vulnerabilities (e.g., CVE, CWE), and indicators of risk (e.g., deprecated, unauthorized).

We implement a case-insensitive regular expression (regex) pattern to efficiently match occurrences of these keywords in SO comments. The complete list of security-related keywords used for filtering is included in our replication package \cite{SOSecureData2025}. Our secure knowledge base contains 43,338 answer posts and 38,827,772 comments for Python, and 2,000 answer posts and 1,467,317 comments for C. Our focus on comments rather than answers is deliberate, as comments often contain critical security insights from the community regarding the proposed solutions. These comments frequently highlight overlooked vulnerabilities or security considerations in otherwise functional code. 

Using this approach, we construct a security-aware knowledge base where each entry is an answer with its corresponding comments, at least one of which contains security-related keywords. This knowledge base effectively serves as an ``antipattern'' repository, capturing collective community insights by documenting instances where members identified potential security concerns.

\subsection{Retrieval System}
Given a code snippet, we retrieve similar answers from the security-aware knowledge base using BM25 \cite{bm25} as an information retrieval system. BM25 is a bag-of-words model that looks for how many of the query terms are present in a document. It ranks the document with the highest number of query terms, normalized by document length, at the top.

We selected BM25 because previous research \cite{rahman2019toward} has demonstrated its effectiveness for software engineering
retrieval tasks, showing that it often outperforms alternatives such as VSM\cite{vsm} and LSI\cite{lsi} when
retrieving across heterogeneous artifacts (e.g., bug reports versus source code). This observation
is consistent with our own experiments. We tested dense retrieval approaches, including FAISS\cite{faiss} 
indexing with sentence embeddings and cosine similarity, but BM25 consistently retrieved more
relevant discussions. In practice, BM25's strength in matching exact tokens such as API names, 
library identifiers, and error patterns makes it a better fit for our security-aware knowledge base.

Using BM25 helps identify code that uses similar libraries and constructs as the reference snippet based on lexical matching. In practice, this filtering stage effectively narrows the search space to a subset of candidate code snippets that use similar libraries. We prioritize library-specific matching because vulnerabilities are often library-dependent. Although the code may be semantically similar, the use of a specific library may introduce security vulnerabilities or safety concerns that would not exist in alternative implementations.

We use the code within the answer, concatenating all code blocks, as the basis for the matching rather than the surrounding text. This choice is motivated by the fact that textual explanations in SO answers may not be as authoritative as the code itself, particularly compared to LLM-generated code. Additionally, textual content may exhibit stylistic variations that do not necessarily reflect meaningful differences in functionality. After retrieving the nearest answers, one or more complete answers, along with their associated comment threads, are included as context.

\section{Evaluation}

This section describes our experimental setup for evaluating SOSecure on our baseline systems under test. We implement SOSecure in Python. We make use of gpt-4o-mini as the LLM for all the experiments. We use the default values for all the hyper-parameters for both GPT4 and BM25.

\subsection{Baseline Systems under test}

\begin{figure*}[!t]
\vspace*{-5mm}
\centering
\small

% Row 1
\begin{subfigure}{0.49\textwidth}
\begin{minipage}{\textwidth}
\begin{promptstyled}{(a) GPT4 (Base)}

{\parbox{\dimexpr\linewidth-2\fboxsep}{
\textbf{Base Prompt:} Review the code for security flaws. 
If issues are found, fix them while keeping original functionality.}}
\newline
\textbf{\{code\}}

\end{promptstyled}
\end{minipage}
\end{subfigure}
\hfill
\begin{subfigure}{0.49\textwidth}
\begin{minipage}{\textwidth}
\begin{promptstyled}{(b) GPT4+CWE}
\textbf{Base Prompt}
\newline
\textbf{\{code\}}

\colorbox{blue!10}{\parbox{\dimexpr\linewidth-2\fboxsep}{
CWE: \textbf{\{CWE\_number\}} \\
}}
\end{promptstyled}
\end{minipage}
\end{subfigure}

\vspace{2mm}

% Row 2
\begin{subfigure}{0.49\textwidth}
\begin{minipage}{\textwidth}
\begin{promptstyled}{(c) GPT4+CWE+}
\textbf{Base Prompt}
\newline
\textbf{\{code\}}

\colorbox{green!10}{\parbox{\dimexpr\linewidth-2\fboxsep}{
CWE: \textbf{\{CWE\_number\}} \\
Description: \textbf{\{CWE\_description\}} \\[0.5ex]
}}
\end{promptstyled}
\end{minipage}
\end{subfigure}
\hfill
\begin{subfigure}{0.49\textwidth}
\begin{minipage}{\textwidth}
\begin{promptstyled}{(d) SOSecure}
\textbf{Base Prompt}
\newline
\textbf{\{code\}}

\colorbox{yellow!15}{\parbox{\dimexpr\linewidth-2\fboxsep}{
Retrieved related SO context:
\newline
\textbf{\{retrieved\_SO\_context\}} \\[0.5ex]
}}
\end{promptstyled}
\end{minipage}
\end{subfigure}
\vspace{-1mm}
\caption{Prompt templates for baseline approaches and SOSecure: 
(a) GPT4 base (no additional context), 
(b) GPT4+CWE (adds CWE number), 
(c) GPT4+CWE+ (adds CWE number and description), 
(d) SOSecure (adds retrieved SO context). 
Colored boxes highlight the additional context each variant provides. 
The complete prompt templates are available in our replication package~\cite{SOSecureData2025}.}
\vspace{-5mm}
\label{fig:prompt-templates}
\end{figure*}
In our study, we compare SOSecure with several baseline systems to assess its effectiveness in
identifying and mitigating security vulnerabilities. The selected baselines encompass a range of 
approaches, from prompting-only methods (GPT4, GPT4+CWE, GPT4+CWE+), to security-focused 
frameworks (SALLM framework\cite{sallm}), to a SOTA LLM-based repair system (VulRepair\cite{vulrepair}). Although other repair
methods exist as surveyed by \citeauthor{zhou2025large}\cite{zhou2025large}, we focus on these representative
baselines because SOSecure is designed as a lightweight post-hoc retrieval layer that complements
rather than replaces specialized repair models. Together, these baselines provide a balanced 
spectrum against which to evaluate SOSecure.

\textbf{\textit{Language Model (GPT4).}}
This baseline utilizes a standard language model GPT4, without any security-specific prompts. The model is provided with code snippets without additional context. This approach serves as a control group, allowing us to assess the inherent capabilities of the language model in generating code without explicit security guidance.

\textbf{\textit{Language Model with Security Prompt (GPT4+CWE).}} All the datasets we use have code snippets mapped to certain  CWE vulnerabilities.
In this scenario, the language model receives prompts that include the name of a specific CWE vulnerability the given code was flagged with. For example, the prompt might instruct the model to``Does this code have any security vulnerabilities such as CWE-079?'' This setup evaluates the model's ability to apply known security concepts to code modification tasks.

\textbf{\textit{Language Model with Additional Security Prompt (GPT4+ CWE+).}}
Building upon the previous baseline, this approach provides the language model with both the CWE name and a description of the vulnerability. By illustrating the vulnerability with a description, we aim to determine two things - whether more context about the CWE and an increased input token length enhance the model's understanding and ability to generate appropriate fixes. Figure~\ref{fig:prompt-templates} illustrates the prompt templates used for 
GPT4, GPT4+CWE, GPT4+CWE+, and SOSecure.

% \paragraph*{\textit{Language Model with StackOverflow Context}}
% Here, the language model is supplied with relevant StackOverflow answers and comments related to the identified security issue. This context leverages community-driven knowledge and discussions to inform the model's response, assessing the impact of incorporating real-world developer insights into the fix generation process.

\textbf{\textit{SALLM framework.}}
The SALLM framework provides a security-centric evaluation of LLMs. It introduces the SALLM dataset (discussed in Section \ref{datasets}), a rule-based repair component, and novel security metrics (secure@k and vulnerable@k, discussed in more detail in Section \ref{metrics}).

\textbf{\textit{VulRepair.}}  VulRepair is a SOTA LLM-based vulnerability repair system that attempts 
to automatically patch insecure code when prompted with vulnerability descriptions. While the 
underlying CodeT5 model is multilingual, the VulRepair framework and its released models have 
been trained and evaluated exclusively on C/C++ vulnerability datasets. For this reason, we use the LLMSecEval 
benchmarks with C code when comparing against VulRepair in our study.

By systematically evaluating SOSecure against these diverse baselines, we aim to gain comprehensive insights into its performance, identify areas for improvement, and validate its utility in enhancing code security.

\subsection{Benchmark datasets}
\label{datasets}
Our first step is to thoroughly review existing vulnerability datasets \cite{llmseceval,pearce2025asleep,sven,sallm,securityeval,franc,lmsys,crossvul} to select base datasets for further investigation. We exclude datasets in \cite{zhou2019devign, li2018vuldeepecker, chakraborty2021deep} as they target a limited set of (2 or 4) projects or vulnerabilities and lack coverage of many typical code generation queries. We found two datasets that have high coverage of CWEs and come with CodeQL files that can be used to detect these in repaired code. We also use LMSys, a dataset based on real-world LLM prompts and responses.

% \begin{figure*}[!t]
%     \centering
%     \includegraphics[width=\columnwidth]{images/prompts.pdf}
%    \vspace{-5mm}
%     \caption{Sample prompt for +CWE} 
%   \vspace{-5mm}
%     \label{fig:prompt}
% \end{figure*}

% \begin{figure*}[!t]
% \vspace*{-5mm} 
% \centering
% \begin{minipage}{\textwidth}
% \begin{promptstyled}{Prompt Template: GPT4+CWE}
% You previously generated the following code:

% \textbf{\{code\}}

% Does this code have any security vulnerabilities such as \textbf{\{CWE\_number\}}?

% Instructions:
% \begin{itemize}[leftmargin=*, itemsep=0.2em, label=-]
%   \item Review the code for any security flaws.
%   \item If security issues are found, modify the code minimally to fix the issue
%         while maintaining its original functionality and logic.
%   \item If no security issues are found, output ``No security issues found''.
% \end{itemize}
% \end{promptstyled}
% \end{minipage}

% \caption{Sample prompt for GPT4{+}CWE}
% \label{fig:prompt}

% \end{figure*}

\begingroup

\setlength{\tabcolsep}{1pt}
\begin{table*}[!t]

\small
\begin{minipage}[t]{0.33\textwidth}
    \centering
    \normalsize 
    \begin{tabular}{@{}lccc@{}}
    \hline
    \rowcolor{gray!25} 
    \textbf{System} & \textbf{FR} & \textbf{IR} & \textbf{NCR} \\
    \hline
   SOSecure & 71.7 & 0 & 48.7 \\
    \rowcolor{gray!8}
    GPT4    & 49.1 & 0 & 64.9 \\
    GPT4+CWE  & 58.5 & 0 & 58.1 \\
    \rowcolor{gray!8}
    GPT4+CWE+  & 60.4  &0  &56.8  \\
    
    \hline
    \end{tabular}
    \subcaption{SALLM dataset}
    \label{tab:sallm}
\end{minipage}%
\begin{minipage}[t]{0.33\textwidth}
    \centering
    \begin{tabular}{@{}lccc@{}}
    \hline
    \rowcolor{gray!25} 
    \textbf{System} & \textbf{FR} & \textbf{IR} & \textbf{NCR} \\
    \hline
    SOSecure & 91.3 & 0 & 57.1 \\
    \rowcolor{gray!8}
    GPT4    & 56.5 & 0 & 73.5 \\
    GPT4+CWE  & 69.6 & 7.7 & 63.3 \\
    \rowcolor{gray!8}
    GPT4+CWE+ & 69.6 & 3.9  & 65.3  \\
    
    \hline
    \end{tabular}
    \subcaption{LLMSecEval}
    \label{tab:llmseceval}
\end{minipage}%
\begin{minipage}[t]{0.33\textwidth}
    \centering
    \begin{tabular}{@{}lccc@{}}
    \hline
    \rowcolor{gray!25} 
    \textbf{System} & \textbf{FR} & \textbf{IR} & \textbf{NCR} \\
    \hline
    SOSecure & 96.7 & 0 & 3.3 \\
    \rowcolor{gray!8}
    GPT4    & 37.5 & 0 & 62.5 \\
    GPT4+CWE  & 45.8 & 0 & 54.1 \\
    \rowcolor{gray!8}
    GPT4+CWE+ & 63.3  & 0  & 36.7 \\
    
    \hline
    \end{tabular}
    \subcaption{LMSys }
    \label{tab:lmsys}
\end{minipage}
\vspace{-1mm}
\caption{Security Metrics Comparison Across Benchmark Datasets. FR: Fix Rate; IR: Introduced vulnerabilities Rate; NCR: No Change Rate.}
\vspace{-18mm}
\label{tab:security-metrics}
\end{table*}

\textit{ \textbf{SALLM dataset}} contains 100 prompts, available in both text and code formats, along with the corresponding generated code snippets covering 45 vulnerability types (CWEs). Each snippet is mapped to a CWE. We select samples that include default CodeQL QL files, resulting in a final set of \textbf{74} samples used for analysis.

\textit{\textbf{LLMSecEval}} is a natural language prompt-to-code dataset crafted from \citeauthor{pearce2025asleep} \cite{pearce2025asleep}. LLMSecEval has 150 prompts instructing an LLM to generate C code(67 samples) and Python code (83 samples). Of the total samples, we choose samples that are flagged with CWEs that include default CodeQL .ql files, resulting in a final set of \textbf{49} samples used for analysis for Python and \textbf{40} for C.

\textit{\textbf{LMSYS-Chat-1M}} consists of 1 million samples, of which 43,269 contain Python code. Among these, 31,008 samples are single-round user conversations with a single code block. To curate a high-quality dataset containing genuine vulnerabilities, we apply a two-step filtration process. First, we run the samples through two static analyzers (Bandit \cite{bandit} and CodeQL) and retain only those flagged as vulnerable by both. This results in 2,809 samples. We further select samples associated with CWEs that include default CodeQL QL files, yielding a final set of \textbf{240} Python samples.

\subsection{Evaluation Metrics}
\label{metrics}

To determine the presence or absence of vulnerabilities in the generated code, we use CodeQL for evaluation. CodeQL is a static analysis tool designed to automatically detect vulnerabilities by executing QL queries on a database generated from the source code. We also study two programming languages, Python and C. This served as an additional motivating factor for using CodeQL, as it allows for a uniform analysis, enabling the use of a single tool to assess both languages. We compute the following security metrics to evaluate SOSecure:

\textbf{\textit{Fix Rate (FR).}} \% of vulnerabilities that were fixed by the system. A higher FR indicates better security performance.

\textbf{\textit{Intro Rate (IR).}} \% of new vulnerabilities introduced after code generation. A lower IR reflects fewer new vulnerabilities.

\textbf{\textit{No-Change Rate (NCR).}} \% of issues that remained unchanged after the code generation. A higher NCR may suggest ineffective issue resolution by the system.

\textbf{\textit{Fixed/Total (F/T).}} The raw count of vulnerabilities successfully repaired relative to the total number of vulnerable samples in the dataset.

\textbf{\textit{Precision Percentage (P\%).}} The proportion of attempted fixes that are correct. A higher P\% indicates better
security performance.

\textbf{\textit{secure@k, vulnerable@k \cite{sallm}}} measures the security of generated code. \textit{vulnerable@k} measures the probability that at least one code snippet out of \textit{k} generated samples is vulnerable. For this metric, a lower score indicates better performance of the system.
The \textit{secure@k} metric measures the probability that all code snippets out of \textit{k} samples are vulnerability-free.  For this metric, a higher score indicates better performance of the system. We use both these metrics to compare the performance of SOSecure and SALLM framework.

\section{Results}
\begin{table}[!t]
    \centering
    \small
    % Requires \usepackage{multirow} and \usepackage{colortbl}
    \begin{tabular}{@{}lccc@{}}
        \hline
        \rowcolor{gray!25} 
        \textbf{Temperature} & \textbf{Metric} & \textbf{SOSecure (\%)} & \textbf{SALLM (\%)} \\
        \hline
        \multirow{2}{*}{\cellcolor{gray!8}0.0} & \cellcolor{gray!8}secure@1 & \cellcolor{gray!8}\textbf{89.19} & \cellcolor{gray!8}51.35 \\
        & \cellcolor{gray!8}vulnerable@1 & \cellcolor{gray!8}\textbf{10.81} & \cellcolor{gray!8}48.65 \\
        \hline
        \multirow{2}{*}{0.2} & secure@1 & 85.14 & 50.0 \\
        & vulnerable@1 & 14.86 & 50.0 \\
        \hline
        \multirow{2}{*}{\cellcolor{gray!8}0.4} & \cellcolor{gray!8}secure@1 & \cellcolor{gray!8}83.78 & \cellcolor{gray!8}51.35 \\
        & \cellcolor{gray!8}vulnerable@1 & \cellcolor{gray!8}16.22 & \cellcolor{gray!8}48.65 \\
        \hline
        \multirow{2}{*}{0.6} & secure@1 & 79.73 & 51.35 \\
        & vulnerable@1 & 20.27 & 48.65 \\
        \hline
        \multirow{2}{*}{\cellcolor{gray!8}0.8} & \cellcolor{gray!8}secure@1 & \cellcolor{gray!8}85.14 & \cellcolor{gray!8}52.70 \\
        & \cellcolor{gray!8}vulnerable@1 & \cellcolor{gray!8}14.86 & \cellcolor{gray!8}47.30 \\
        \hline
        \multirow{2}{*}{1.0} & secure@1 & 75.68 & 52.70 \\
        & vulnerable@1 & 24.32 & 47.30 \\
        \hline
    \end{tabular}

    \caption{Comparison of SOSecure with SALLM framework}
    \vspace{-10mm}
    \label{tab:sosecure-vs-sallm}
\end{table}

In this section, we present the evaluation results of SOSecure in terms of its effectiveness, generalizability, CWE types, and neighbor sensitivity across multiple datasets, and compare its performance with baseline approaches. We analyze the effectiveness of our approach in addressing security vulnerabilities in LLM-generated code and examine its performance across different CWEs and programming languages.

\subsection{Effectiveness of SOSecure}
Table \ref{tab:security-metrics} presents the comparison of SOSecure with baseline approaches across three benchmark datasets. The results demonstrate that SOSecure consistently outperforms standard LLM approaches in terms of security vulnerability mitigation.
On the SALLM dataset, SOSecure achieves a Fix Rate (FR) of 71.7\%, significantly higher than both GPT4 (49.06\%) and GPT4+CWE (58.49\%). Similarly SOSecure achieves a 91.3\% Fix Rate compared to 56.52\% for GPT4 and 69.57\% for GPT4+CWE. The most substantial performance gain is observed on the LMSys dataset, where SOSecure achieves a remarkable 96.67\% Fix Rate, compared to only 37.50\% for GPT4 and 45.83\% for GPT4+CWE.
Importantly, while improving security, SOSecure does not introduce new vulnerabilities, maintaining a 0\% Introduced vulnerabilities Rate (IR) across all datasets. In contrast, GPT4+CWE introduces new vulnerabilities in 7.69\% of cases on the LLMSecEval dataset. The No Change Rate (NCR) is also consistently lower for SOSecure (48.65\%, 57.14\%, and 3.33\% across the three datasets) compared to both baseline approaches.

Beyond GPT4 variants, we also compare SOSecure against two systems designed specifically
for secure code generation and repair: the SALLM framework and VulRepair. 

Table~\ref{tab:sosecure-vs-sallm} shows that the SALLM framework baseline, 
as reported in their paper~\cite{sallm}, achieves roughly 51\% secure@1, 
whereas SOSecure consistently attains 75--89\% secure@1 for varying temperatures, 
nearly doubling the proportion of secure generations. Vulnerable@1 also drops
from about 49\% with SALLM to as low as 10--15\% with SOSecure.

Table~\ref{tab:llmseceval_combined} reports that VulRepair achieves only a 13.3\% Fix Rate 
with a 95\% No-Change Rate, leaving most vulnerabilities unaddressed. In contrast, SOSecure fixes 73.3\% of 
vulnerabilities while introducing none (0\% IR). 

Overall, SOSecure outperforms all categories of baselines across datasets: it achieves higher fix rates than prompting-only techniques (GPT4, GPT4+CWE, GPT4+CWE+), stronger secure@k and lower vulnerable@k scores than the security-centric SALLM framework, and surpasses the repair system VulRepair in fix rate on the LMSys C subset. These results highlight the value of community-driven retrieval and demonstrate SOSecure’s effectiveness as a lightweight post-hoc security layer for LLM code generation.

\begin{table*}[!t]
    \centering
    \normalsize 
    \begin{subfigure}[b]{0.48\textwidth}
        \centering
        \small
        \begin{tabular}{l c cc cc cc cc}
            \toprule
            \rowcolor{gray!15} 
            & & \multicolumn{2}{c}{\textbf{SOSecure}}
    & \multicolumn{2}{c}{\textbf{GPT4}}
    & \multicolumn{2}{c}{\textbf{GPT4+CWE}}
    & \multicolumn{2}{c}{\textbf{VulRepair}} \\
  \cmidrule(lr){3-4}\cmidrule(lr){5-6}\cmidrule(lr){7-8}\cmidrule(lr){9-10}
  \rowcolor{gray!15}
  \textbf{CWE} & \textbf{Sev}
    & \textbf{F/T} & \textbf{P\%}
    & \textbf{F/T} & \textbf{P\%}
    & \textbf{F/T} & \textbf{P\%}
    & \textbf{F/T} & \textbf{P\%}   \\
  \midrule
            
            078   & 9.8 & 6/6 & 100 & 2/6 & 33.3 & 3/6 & 50 & 2/6 & 33.3\\
            
            \rowcolor{gray!5} 190  & 8.6 & 2/2 & 100 & 2/2 & 100 & 2/2 & 100 & 0/2 
            & 0\\
            
             022 & 7.5 & 0/3 & 0 & 0/3 & 0 & 0/3 & 0 & 0/3 & 0\\
            
            \midrule
            \rowcolor{gray!15} \textbf{Avg} &  & & \textbf{66.7} & & \textbf{44.4} & & \textbf{50} & &
            \textbf{11.1}\\
            \bottomrule
        \end{tabular}
        \subcaption{}
        \label{tab:performance_llmseceval_c}
    \end{subfigure}
    \hfill
    \begin{subfigure}[b]{0.48\textwidth}
        \centering
        \normalsize 
        \begin{tabular}{@{}lccc@{}}
            \hline
            \rowcolor{gray!25} 
            \textbf{System} & \textbf{FR (\%)} & \textbf{IR (\%)} & \textbf{NCR (\%)} \\
            \hline
           SOSecure & 73.3 & 0 & 72.5 \\
            \rowcolor{gray!8}
            GPT4    & 53.3 & 0 & 80 \\
            GPT4+CWE  & 60 & 0 & 77.5 \\
            \rowcolor{gray!8}
            GPT4+CWE+  & 53.3 & 0 & 80 \\
           \rowcolor{gray!8}
            VulRepair  & 13.3 & 0 & 95 \\

            \hline
        \end{tabular}
        \subcaption{}
        \label{tab:llmsecevalC}
    \end{subfigure}
    \caption{Performance of SOSecure on C code from the LLMSecEval dataset. The left side shows vulnerability mitigation results by CWE type, with F/T representing fixed/total vulnerabilities and P\% showing precision percentage. The right side presents system-level metrics comparing SOSecure against baseline approaches, showing Fix Rate (FR\%), Introduced vulnerability Rate (IR\%), and No Change Rate (NCR\%).}
    \vspace{-8mm}
  
    \label{tab:llmseceval_combined}
    
\end{table*}
\subsection{Performance By Vulnerability Type}
Table \ref{tab:performance_all} provides a detailed breakdown of performance by CWE type across the three datasets for CWEs listed as the Top 25 Most Dangerous Software Weaknesses.\footnote{\url{https://cwe.mitre.org/top25/}} The results reveal that SOSecure demonstrates varying effectiveness depending on the vulnerability type, with particularly strong performance on high-severity and critical vulnerabilities according to the CVSS scoring system.\footnote{Severity levels based on CVSS scores as defined by GitHub CodeQL: \url{https://github.blog/changelog/2021-07-19-codeql-code-scanning-new-severity-levels-for-security-alerts/}}

In the SALLM dataset, SOSecure achieves perfect precision (100\%) for CWE-918 (Server-Side Request Forgery) and CWE-089 (SQL Injection), which are critical (9.1) and high (8.8) severity vulnerabilities, respectively. For CWE-094 (Code Injection, severity 9.3), a critical vulnerability, SOSecure achieves 77.78\% precision, outperforming GPT4 (55.56\%) while matching GPT4+CWE. SOSecure also demonstrates superior performance for CWE-078 (OS Command Injection, severity 6.3), a medium severity vulnerability, achieving 90\% precision compared to 80\% for both baseline approaches.

Similar patterns are observed in the LMSys dataset, where SOSecure achieves 100\% precision for CWE-094 and strong performance across other vulnerability types. In the LLMSecEval dataset, SOSecure achieves perfect precision (100\%) for five out of seven CWE types, including the critical vulnerabilities CWE-502 (Deserialization of Untrusted Data, severity 9.8) and CWE-798 (Use of Hard-coded Credentials, severity 9.8).

Notably, SOSecure demonstrates consistent improvement across vulnerability types of varying severity levels, indicating that the Stack Overflow discussions provide valuable security insights across a broad spectrum of security concerns. The most substantial improvements are observed for critical (CVSS 9.0-10.0) and high (CVSS 7.0-8.9) severity vulnerabilities, suggesting that community discussions are particularly valuable for addressing the most dangerous security weaknesses.

\subsection{Impact of Number of Retrieved Candidates}

Figure \ref{fig:knn} illustrates the effect of varying the number of neighbors (k) added as context on the average precision of SOSecure for LMSys data. We observe that performance improves when increasing from 1 to 3 neighbors, after which it plateaus and declines slightly beyond 5 neighbors. This suggests that excessive context may introduce noise or conflicting information, whereas too few discussions may not provide sufficient security insights. An intermediate number of neighbors therefore provides the best balance, maintaining focus on the specific vulnerability while still providing useful corrective context.

\begin{table*}[t]
    \centering
    
    \begin{subfigure}[b]{0.31\textwidth}
        \centering
        \large
        \resizebox{\textwidth}{!}{%
        \begin{tabular}{l c cc cc cc}
            \toprule
            \rowcolor{gray!15} 
            & & \multicolumn{2}{c}{\textbf{SOSecure}} & \multicolumn{2}{c}{\textbf{GPT4}} & \multicolumn{2}{c}{\textbf{GPT4+CWE}} \\
            \cmidrule(lr){3-4} \cmidrule(lr){5-6} \cmidrule(lr){7-8}
            \rowcolor{gray!15} 
            \textbf{CWE} & \textbf{Sev} & \textbf{F/T} & \textbf{P\%} & \textbf{F/T} & \textbf{P\%} & \textbf{F/T} & \textbf{P\%} \\
            \midrule
            094   & 9.3 & 9/9 & 77.8 & 5/9 & 55.6 & 7/9 & 77.8 \\
            \rowcolor{gray!5} 918  & 9.1 & 1/1 & 100 & 1/1 & 100 & 1/1 & 100\\
             089  & 8.8 & 2/2 & 100 & 2/2 & 100 & 2/2 & 100 \\
             \rowcolor{gray!5}020  & 7.8 & 2/4 & 25 & 1/4 & 25 & 1/4 & 25 \\
             022  & 7.5 & 0/2 & 0 & 0/2 & 0 & 0/2 & 0 \\
            \rowcolor{gray!5} 078  & 6.3 & 9/10 & 90 & 8/10 & 80 & 9/10 & 90 \\
            079  & 6.1 & 3/4 & 75 & 3/4 & 75 & 3/4 & 75 \\
            \midrule
            \rowcolor{gray!15} \textbf{Avg} &  & & \textbf{73.6} & & \textbf{62.2} & & \textbf{66.8} \\
            \bottomrule
        \end{tabular}%
        }
        \caption{SALLM}
        \label{tab:performance_sallm}
    \end{subfigure}
    \hfill
    \begin{subfigure}[b]{0.31\textwidth}
        \centering
        \large 
        \resizebox{\textwidth}{!}{%
        \begin{tabular}{l c cc cc cc}
            \toprule
            \rowcolor{gray!15} 
            & & \multicolumn{2}{c}{\textbf{SOSecure}} & \multicolumn{2}{c}{\textbf{GPT4}} & \multicolumn{2}{c}{\textbf{GPT4+CWE}} \\
            \cmidrule(lr){3-4} \cmidrule(lr){5-6} \cmidrule(lr){7-8}
            \rowcolor{gray!15} 
            \textbf{CWE} & \textbf{Sev} & \textbf{F/T} & \textbf{P\%} & \textbf{F/T} & \textbf{P\%} & \textbf{F/T} & \textbf{P\%} \\
            \midrule
            502   & 9.8 & 8/8 & 77.78 & 5/8 & 62.5 & 8/8 & 100 \\
            \rowcolor{gray!5} 094  & 9.3 & 188/188 & 100 & 68/188 & 36.2 & 66/188 & 35.10\\
             022  & 7.5 & 1/1 & 100 & 1/1 & 100 & 1/1 & 100 \\
             \rowcolor{gray!5}078  & 6.3 & 6/7 & 85.7 & 5/7 & 71.4 & 7/7 & 100 \\
            \midrule
            \rowcolor{gray!15} \textbf{Avg} &  & & \textbf{96.4} & & \textbf{67.5} & & \textbf{83.8} \\
            \bottomrule
        \end{tabular}%
        }
        \caption{LMSYS}
        \label{tab:performance_lmsys}
    \end{subfigure}
    \hfill
    \begin{subfigure}[b]{0.31\textwidth}
        \centering
        \large 
        \resizebox{\textwidth}{!}{%
        \begin{tabular}{l c cc cc cc}
            \toprule
            \rowcolor{gray!15} 
            & & \multicolumn{2}{c}{\textbf{SOSecure}} & \multicolumn{2}{c}{\textbf{GPT4}} & \multicolumn{2}{c}{\textbf{GPT4+CWE}} \\
            \cmidrule(lr){3-4} \cmidrule(lr){5-6} \cmidrule(lr){7-8}
            \rowcolor{gray!15} 
            \textbf{CWE} & \textbf{Sev} & \textbf{F/T} & \textbf{P\%} & \textbf{F/T} & \textbf{P\%} & \textbf{F/T} & \textbf{P\%} \\
            \midrule
            502   & 9.8 & 4/4 & 100 & 4/4 & 100 & 4/4 & 100 \\
            \rowcolor{gray!5} 798  & 9.8 & 3/3 & 100 & 2/3 & 66.7 & 2/3 & 66.7\\
             089  & 8.8 & 5/5 & 100 & 5/5 & 100 & 5/5 & 100 \\
             \rowcolor{gray!5}020  & 7.8 & 2/2 & 100 & 0/2 & 0 & 2/2 & 100 \\
             022  & 7.5 & 4/6 & 66.7 & 0/6 & 0 & 1/6 & 16.7 \\
             \rowcolor{gray!5}078  & 6.3 & 2/2 & 100 & 1/2 & 50 & 1/2 & 50 \\
             079  & 6.1 & 1/1 & 100 & 1/1 & 100 & 1/1 & 100 \\
            \midrule
            \rowcolor{gray!15} \textbf{Avg} &  & & \textbf{95.2} & & \textbf{59.5} & & \textbf{76.2} \\
            \bottomrule
        \end{tabular}%
        }
        \caption{LLMSecEval}
        \label{tab:performance_llmsecevalsub}
    \end{subfigure}
    \caption{Performance Evaluation of 2024 CWE Top 25 Most Dangerous Software Weaknesses across three datasets. Metrics shown: Severity CodeQL (Sev), Fixed/Total vulnerabilities (F/T), and Precision\% (P\%).}
    \vspace{-10mm}
   
    \label{tab:performance_all}
\end{table*}

\subsection{Performance Across Languages}
To evaluate the language generalizability of our approach, we tested SOSecure on both Python and C code samples from the LLMSecEval dataset. Results are presented in Table \ref{tab:llmseceval_combined}.
SOSecure maintained strong performance across both languages, achieving a Fix Rate (FR) of 73.33\% on C code samples compared to 53.33\% for GPT4 and 60\% for GPT4+CWE. Similar to our findings with Python, SOSecure maintained a 0\% Introduced vulnerabilities Rate (IR) on C code.
When analyzed by CWE type for C code, SOSecure achieved 100\% precision for CWE-078 (OS Command Injection) and CWE-190 (Integer Overflow), demonstrating that our approach generalizes well across different programming languages. However, SOSecure showed limitations with CWE-022 (Path Traversal) in C code, similar to our findings with Python.
These results suggest that SOSecure's approach of leveraging SO discussions as context is effective across multiple programming languages.

\section{Discussion}

Our findings align with recent work showing that RAG approaches can effectively inject time-sensitive and domain-specific knowledge into general-purpose LLMs \cite{vulrag}. In the security context, this injection is particularly valuable given the rapid evolution of security best-practices and the discovery of new vulnerability patterns. Our experimental results demonstrate that SOSecure enhances the security of LLM-generated code by leveraging community knowledge from SO. In this section, we discuss the broader implications of our findings, analyze the strengths and limitations of our approach, and consider its practical applications.

\subsection{Qualitative Analysis of Success Cases}

To better understand how SOSecure improves code security, we conducted a qualitative analysis of both successful fixes and failure cases. We selected annotations from the LMSys dataset to evaluate performance with realistic, user-generated queries that represent actual code generation scenarios encountered in real-world usage. This analysis provides insights into the mechanisms by which community knowledge contributes to security improvements, as well as the limitations of our approach.

\subsubsection{Applications of Community Security Insights} ~\\ 

\noindent
\textbf{Case Study 1: Command Injection Vulnerability (CWE-078)}
\noindent
In this example, SOSecure successfully mitigated a command injection vulnerability in a Flask application:
\begin{lstlisting}[language=Python, caption=Original vulnerable code]
@app.route('/execute')
def executeCommand():
    command = request.args.get('command')
    subprocess.call(command, shell=True)
    return command
\end{lstlisting}

\noindent
The SO comment directly addressing this issue stated: \textit{``Don't use shell=True to run commands, it opens the program to command injection vulnerabilities."}

\noindent
This explicit security warning from the community was directly applied in the fixed implementation:

\begin{lstlisting}[language=Python, caption=Fixed code with SOSecure]
@app.route('/execute')
def executeCommand():
    command_key = request.args.get('command')
    if command_key not in ALLOWED_COMMANDS:
        abort(403) # Forbidden
    command = ALLOWED_COMMANDS[command_key]
    result = subprocess.check_output(command.split(), stderr=subprocess.STDOUT)
    return result.decode()
\end{lstlisting}

\noindent
The fix implements exactly what the SO comment suggests: avoiding \texttt{shell=True} and using a command whitelist approach, demonstrating a direct translation of community knowledge into security improvements.
\begin{figure*}[!t]

    \centering
    
    \includegraphics[width=0.5\columnwidth]{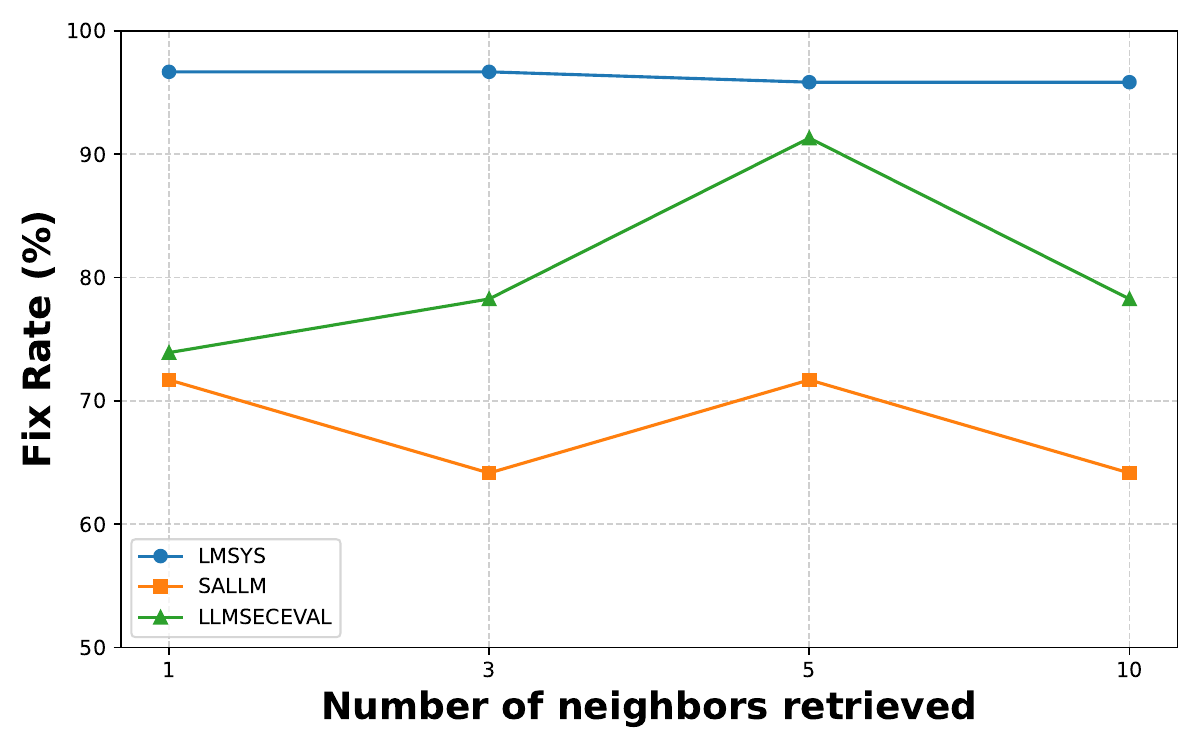}
   \vspace{-3mm}
    \caption{Effect of the number of neighbors added as context on the Fix Rate\%.}
    \vspace{-3mm}
    \label{fig:knn}
\end{figure*}
~\\ \noindent
~\\
\textbf{Case Study 2: XML External Entity (XXE) Vulnerability (CWE-611)}
\noindent
For an XXE vulnerability in XML parsing:

\begin{lstlisting}[language=Python, caption=Original vulnerable code]
@app.route('/parse_xml')
def parse_xml():
    xml_src = request.args.get('xml_src').encode('utf-8')
    doc = lxml.etree.fromstring(xml_src)
    return lxml.etree.tostring(doc)
\end{lstlisting}

\noindent
While no SO comment directly mentioned XXE, the community discussion contextualized proper Flask request handling, which SOSecure combined with security best practices generating this fix:
~\\
\begin{lstlisting}[language=Python, caption=Fixed code with SOSecure]
@app.route('/parse_xml')
def parse_xml():
    xml_src = request.args.get('xml_src')
    if not xml_src:
        return Response("Missing 'xml_src' parameter.", status=400)
    try:
        # Parse the XML while disabling external entity processing
        parser = lxml.etree.XMLParser(resolve_entities=False)
        doc = lxml.etree.fromstring(xml_src.encode('utf-8'), parser)
        return lxml.etree.tostring(doc)
    except lxml.etree.XMLSyntaxError:
        return Response("Invalid XML provided.", status=400)
\end{lstlisting}

\noindent
\textbf{Case Study 3: Debug Mode Security Vulnerability (CWE-094)}

\noindent
A particularly clear example of direct knowledge transfer involves fixing Flask's debug mode vulnerability. In multiple instances, SOSecure encountered code running Flask in debug mode:

\begin{lstlisting}[language=Python]
app.run(debug=True)
\end{lstlisting}

\noindent
The associated SO comments implicitly mentioned security concerns, such as:
\textit{``the last lines are good habit for projects like these, debugger on is extremely useful."}

\noindent
SOSecure correctly identified this as security advice intended only for development environments, and consistently made the appropriate fix:

\begin{lstlisting}[language=Python, caption=Fixed code with SOSecure, breaklines=true]
app.run(debug=False)  # Set debug to False for production
\end{lstlisting}

\noindent
or even better:

\begin{lstlisting}[language=Python, caption=Environment-aware fix, breaklines=true]
debug_mode = os.environ.get('FLASK_DEBUG', '0') == '1'
app.run(debug=debug_mode)
\end{lstlisting}

\noindent
This pattern appeared in numerous examples, showing SOSecure's ability to recognize when developer-focused advice needs to be adapted for production security.

~\\ \noindent
\textbf{Case Study 4: Code Injection Risk (CWE-94)}

\noindent
An example of more indirect knowledge transfer relates to dangerous serialization methods. For the following listing:

\begin{lstlisting}[language=Python, caption=Original vulnerable code, breaklines=true]
r.set("test", pickle.dumps(request.json))
result = pickle.loads(r.get("test"))

\end{lstlisting}

\noindent
The retrieved SO comment\footnote{\url{https://stackoverflow.com/a/35580819}} warned: \textit{``never ever ever use `eval' in a web application, way too many attack vectors there..."}

\noindent
Followed by another comment responding: \textit{``fair enough. edited to use `pickle' instead of `eval'"}

\noindent
Although the input code already used \texttt{pickle}, SOSecure correctly recalled that \texttt{pickle} also presents security risks and transformed the code to use safer JSON serialization:

\begin{lstlisting}[language=Python, caption=Fixed code with SOSecure, breaklines=true]
r.set("test", json.dumps(request.json))
result = json.loads(r.get("test"))

\end{lstlisting}

\noindent GPT4, when prompted to consider repairing this snippet, did not make this change.

\subsubsection{Patterns in Effective Community Knowledge Transfer}
Across the analyzed examples, several patterns emerged in how SOSecure effectively leverages community knowledge (when backed by a capable LLM). When SO comments explicitly mention security concerns (e.g., "Don't use shell=True"), SOSecure directly applies these insights in its fixes. Even when comments don't explicitly mention a vulnerability type, they often provide contextual information about proper framework usage that SOSecure combines with security best practices. SOSecure does not blindly apply all suggestions from SO, but critically evaluates them for security implications, as seen in the pickle/eval example. Many effective fixes leverage community discussions about framework-specific practices, such as Flask's debug mode, demonstrating how framework expertise contributes to security. These patterns highlight the various mechanisms through which community knowledge enhances the security of LLM-generated code.

\subsection{Qualitative Analysis of Failure Cases}

\noindent
\textbf{Case Study 1: Improper SSL/TLS Configuration (CWE-327)}

\noindent
Here, SOSecure attempted to fix a vulnerable SSL implementation:

\begin{lstlisting}[language=Python, caption=Original vulnerable code]
# Original vulnerable code
context = ssl.create_default_context()
context.check_hostname = True
context.verify_mode = ssl.CERT_REQUIRED

try:
    # Assume recipient is a valid hostname
    with socket.create_connection((recipient, 443)) as sock:
        with context.wrap_socket(sock, server_hostname=recipient) as ssl_sock:
            ssl_sock.sendall(decrypted_data)
            response = ssl_sock.recv(4096)
\end{lstlisting}

\noindent
The SO comments contained valuable but indirect security advice: \textit{``Since the `Poodle' vulnerability in SSLv3 many webservers have disabled it. You may need to add \\ \texttt{ssl\_version=ssl.PROTOCOL\_TLSv1} to your \\ \texttt{get\_server\_certificate(...)} call''}

\noindent
While SOSecure addressed the original flagged vulnerability (default SSL/TLS version), the fixed code still had this issue:

\begin{lstlisting}[language=Python, caption=Fixed code with SOSecure (still vulnerable)]
# Fixed code with SOSecure (still vulnerable)
context = ssl.create_default_context()
context.check_hostname = True
context.verify_mode = ssl.CERT_REQUIRED
\end{lstlisting}

\noindent
CodeQL still flagged this code with: \textit{``Insecure SSL/TLS protocol version TLSv1 allowed by call to $ssl.create\_default\_context$."}

\noindent
The failure occurred because the SO comment mentioned using TLSv1, which is now considered insecure, but didn't mention that modern applications should use TLSv1.2 or higher. SOSecure did not fully update the SSL configuration to disable older protocols. This illustrates a challenge with evolving security standards: community discussions may not always reflect the most recent security best practices.

\noindent \textbf{Case Study 2: Weak Password Hashing (CWE-327)} \\ \noindent
In another example, SOSecure failed to properly address a weak password hashing implementation:

\begin{lstlisting}[language=Python, caption=Original vulnerable code]
# Original vulnerable code
def getKey(password):
    hasher = SHA256.new(password.encode('utf-8'))
    return hasher.digest()
\end{lstlisting}

\noindent
Although SOSecure made improvements to the encryption and padding mechanisms, it did not address the fundamental issue of using a fast hash function (SHA-256) for password hashing, which was flagged by CodeQL: \textit{``Sensitive data (password) is used in a hashing algorithm (SHA256) that is insecure for password hashing, since it is not a computationally expensive hash function."}

\noindent
The SO comment\footnote{\url{https://stackoverflow.com/a/54197420}} mentioned various encryption issues but didn't specifically address this weakness: \textit{``The general problems I have seen in SO: 1) Encryption modes are incompatible, 2) key sizes are incompatible 3) KDF are not compatible, 4) IV forget 4) output encoding and decoding problems, 5) padding are forgotten, 6) paddings are not compatible,..."}

\noindent
SOSecure should have implemented a more secure password hashing algorithm like bcrypt, Argon2, or PBKDF2 with proper salting and iteration counts. This case illustrates that SOSecure struggles when the necessary security guidance is not explicitly mentioned in the retrieved community discussions.

\noindent \textbf{Case Study 3: Accepting Unknown SSH Host Keys (CWE-295)}

\noindent
In a third example involving SSH connections using Paramiko, SOSecure failed to fully secure the code:

\begin{lstlisting}[language=Python, caption=Original vulnerable code]
# Original vulnerable code
ssh = paramiko.SSHClient()
ssh.set_missing_host_key_policy(paramiko.AutoAddPolicy())
ssh.connect("hostname", username="username", password="password")
\end{lstlisting}

\noindent
SOSecure replaced hardcoded credentials with a prompt, which was an improvement:

\begin{lstlisting}[language=Python, caption=Fixed code with SOSecure (partially fixed)]
# Fixed code with SOSecure (partially fixed)
def connect_sftp(hostname, username):
    password = getpass.getpass(prompt='Enter your password: ')
    ssh = paramiko.SSHClient()
    ssh.set_missing_host_key_policy(paramiko.AutoAddPolicy())
    ssh.connect(hostname, username=username, password=password)
    return paramiko.SFTPClient.from_transport(ssh.get_transport())
\end{lstlisting}

\noindent
However, it failed to address the \texttt{AutoAddPolicy()} issue flagged by CodeQL: \textit{``Setting missing host key policy to AutoAddPolicy may be unsafe."}

\noindent
The SO discussions did not mention this security concern, providing no guidance on proper host key validation for SSH connections.
This failure demonstrates that even when SOSecure can improve some aspects of security (credential handling), it may miss other critical vulnerabilities when community discussions do not address them.

\subsubsection{Patterns in Failure Cases}
Analyzing these and other failure cases reveals several recurring patterns in situations where SOSecure is unable to fully remediate security vulnerabilities. When community discussions refer to outdated security practices that were once considered secure but are now vulnerable (such as TLSv1), SOSecure cannot always discern that these recommendations need to be updated to current standards. SOSecure struggles when secure implementations require knowledge that isn't explicitly stated in the retrieved SO discussions, which is particularly challenging for domain-specific security practices. In many failure cases, SOSecure makes partial improvements to security, such as replacing hardcoded credentials or improving error handling, but misses deeper architectural or protocol-level vulnerabilities that require specialized security expertise. When security decisions involve trade-offs between usability and security that are context-dependent, SOSecure may not have sufficient information to make the optimal decision. We also find that retrieval sometimes returns discussions that focus on debugging rather than security, leading to incomplete fixes.
Future work may be able to address these limitations by sampling multiple, complementary discussions to be used in the prompt rather than just the top-k neighbors. For many patterns, it may even be beneficial to aggregate all security insights from SO discussions into regularly updated knowledge bases to produce shorter and more effective rewrite prompts.

\subsection{Code Similarity Analysis}
Since most of the evaluation datasets do not have test cases, we rely on extensive manual annotation and careful prompting to ensure that SOSecure does not simply remove risky portions of code to address the vulnerability. We did not find evidence of this behavior. We also quantify the average extent of modifications made to the code by SOSecure by calculating the difference between the original and fixed code for the LMSys dataset. We find that successful vulnerability repairs maintained an average similarity of 0.60 (using Python's \texttt{difflib} library, on a scale from 0 to 1), indicating that SOSecure is making targeted security modifications rather than extensive rewrites. We are therefore reasonably confident that it preserves the developer's original intent while addressing specific security concerns, when backed by models at least as capable as GPT4 (used in this work).

\textbf{Token overhead:} Incorporating SO discussions into the prompt context increases the input token count for LLM queries. This additional context comes with increased computational costs and latency, which must be considered for practical deployments. We measure this increase in tokens on the real-world dataset of LMSys with the number of neighbors set to 1 and find that on average, SOSecure adds approximately 530 tokens per query (estimated using tiktoken\footnote{\url{https://github.com/openai/tiktoken}}).

\subsection{The Importance of Community Forums}
The performance of SOSecure across the three data sets validates our core hypothesis that community-driven security insights can meaningfully improve LLM code generation. This effectiveness stems from the key advantages of SO discussions over LLMs: they naturally capture evolving security practices and adapt quickly to newly discovered vulnerabilities. The significant improvement in fix rates for high-severity vulnerabilities (e.g., CWE-094, CWE-502) suggests that SO discussions provide valuable contextual information about security implications specific to particular libraries or API usage patterns, with this contextual awareness evident in cases like CWE-078 (OS Command Injection), where community comments explicitly warn against unsafe practices. Additionally, the collective wisdom captured in SO discussions represents insights from security experts, library maintainers, and experienced developers across various domains, with security knowledge being highly domain specific and vulnerabilities often tied to particular frameworks, libraries, or implementation contexts. This diversity of expertise contributes to SOSecure's ability to address a wide range of vulnerability types effectively.

An important implication of our work is related to the evolving relationship between community knowledge platforms such as SO and AI code generation tools. As developers increasingly rely on LLMs for code generation, forums like SO have seen a steep decrease in usage. There is a risk that fewer new discussions will be created on community platforms, not to mention that more and more of these discussions will be AI-authored, potentially diminishing this valuable knowledge resource. As shown, this would also harm AI-based methods. SOSecure demonstrates the continued value of community discussions as a complement to LLM capabilities.

\subsection{Limitations}

\textbf{CodeQL accuracy constraints.} Our evaluation relies on CodeQL for vulnerability detection, 
which can yield false positives and false negatives. The analysis is further limited to CWEs
covered by CodeQL's default queries, leaving out other potentially important vulnerability classes. 
Consequently, our results capture improvements on vulnerabilities detectable by CodeQL, while
categories such as logic flaws, side channels, or misconfigurations remain outside of its scope. 
Incorporating complementary analysis tools and broader evaluation criteria is a promising direction
for future work.

\noindent
\textbf{Retrieval strategy:} SOSecure's effectiveness depends on the availability and quality of security-related discussions in SO. For newer or niche technologies with limited community discourse, the approach may be less effective. 

Although we experimented with dense retrieval methods (e.g., FAISS with 
cosine similarity), BM25 consistently performed better in our setting, particularly for queries 
involving API names and error messages. We therefore adopt BM25 in SOSecure. Future work may 
revisit semantic or hybrid retrieval approaches that combine lexical precision with semantic recall.

% \section{Limitations and Future Work}
% \input{Limitations}

\section{Conclusion}
This paper introduces SOSecure, a retrieval-augmented approach for improving security in LLM-generated code using community knowledge from Stack Overflow. SOSecure constructs a security-oriented knowledge base from posts and comments containing explicit security warnings, retrieves relevant discussions, and incorporates them as context during code revision.
Our evaluation across three datasets and two languages demonstrates the effectiveness of SOSecure in mitigating security vulnerabilities. 
SOSecure does not require retraining or specialized fine-tuning, allowing seamless integration into existing LLM deployments with minimal overhead. Although the primary evaluation focused on Python, the results from the C language dataset (Table \ref{tab:llmseceval_combined}) show that SOSecure can generalize between programming languages. Additionally, as security discussions evolve on platforms like SO, SOSecure's knowledge base can be continuously updated, ensuring ongoing improvements in security without the need for retraining the model.

\section*{Data Availability}
All data, prompts, and evaluation artifacts are available \href{https://github.com/manishamukherjee/SOSecure}{here} ~\cite{SOSecureData2025}.

\bibliographystyle{ACM-Reference-Format}
\bibliography{references}

\end{document}